# Andreev reflection and Yanson point-contact spectroscopy of a Zn single crystal.

Yu.G. Naidyuk

*B. Verkin Institute for Low Temperature Physics and Engineering, National Academy of Sciences of Ukraine, 47 Lenin Avenue, 61103, Kharkiv, Ukraine*

Kurt Gloos

*Wihuri Physical Laboratory, Department of Physics and Astronomy, University of Turku, FIN-20014 Turku, and Turku University Centre for Materials and Surfaces (MatSurf), FIN-20014 Turku, Finland*

## Abstract

Andreev reflection (AR) and Yanson point-contact spectroscopy (PCS) have been applied simultaneously to study the superconducting (SC) gap and the electron-phonon interaction (EPI) in a Zn single crystal. The correlation between SC gap value and EPI spectrum allowed us to establish the anisotropy of the SC gap. Evidence for multiband superconductivity in Zn is present with two gaps related as 1:1.3. We also found that the AR features are more robust against the point-contact quality than the EPI ones, possibly, due to the large coherence length in Zn compared to the typical PC size. Even for the PCs close to the ballistic regime with intense EPI spectra, the transmission coefficient evaluated from the AR spectra is near the one predicted theoretically for the diffusive regime of a current flow in the PC. Understanding those phenomena would provide a more reliable basis to apply both types of spectroscopies to study more complex SCs.

**Introduction.**

Point-contact spectroscopy (PCS), discovered by I. K. Yanson in 1974 [1], is a powerful tool to investigate the electron-phonon interaction (EPI) in conductors [2]. Electrons, accelerated by applying a voltage *V* at a ballistic point-contact (PC) at low temperatures, release their excess energy e*V* in scattering processes by generating quasi-particles (e.g. phonons). This electron scattering produces a back-flowing current that makes the *I-V* characteristic nonlinear. The second derivative of the *I-V* characteristic represents directly the EPI function as shown theoretically by Kulik, Omelyanchouk and Shekhter in 1977 [3]. During the 1970ths, PCs were also intensively used to study the superconducting (SC) state. In the late of 1970ths, Artemenko, Volkov, and Zaitsev [4] were the first to employ Andreev reflection (AR) to explain the nature of the so-called excess current of the *I-V* characteristics of SC PCs, which had been observed before by different experimentalists (see, e.g., [5] as one of the earliest paper in this field). Moreover, this group also predicted [4] that the differential conductance of a "dirty" PC has a maximum at the SC gap value. Shortly afterwards, this was experimentally observed by their colleagues Gubankov and Margolin [6]. These findings stimulated further theoretical treatments of superconducting PCs so that in 1982 Blonder, Tinkham, and Klapwijk (BTK) published their "simple theory for the *I-V* curves of normal-superconducting microconstriction contacts which describes the crossover from metallic to tunnel junction behavior" [7]. Up to this day, the BTK theory and its modifications are widely used to extract the SC gap Δ and other parameters of superconductors from the experimental *I-V* curves of PCs as well as their derivatives. This is known presently as Andreev reflection spectroscopy. A few years later in 1984, Zaitsev [8] developed a more sophisticated theory beyond the one-dimensional BTK model. It turned out that Zaitsev's final formulae were similar to the BTK equations and produced the same results. Around that time and soon after both theories had been published, heavy-fermion systems and high-temperature superconductors appeared as new classes of superconductors, and experimental efforts were directed to investigate the SC state of these as well as other novel materials using PCs. However, the investigation of classical superconductors is still of interest in order to understand the details of AR phenomena that are found in the study of more complex materials.

Here we examine the conventional superconductor Zn by applying simultaneously Yanson PCS and AR spectroscopy at the same PC. Yanson PCS allows to characterize the metal in the PC core and to determine the current flow regime, whether the contact is ballistic, diffusive, or thermal [2]. Since Zn has a strongly anisotropic PC EPI function [9], a SC gap anisotropy has to be expected. This is what we investigate in this paper using AR spectroscopy.

**Experimental details and results**

The samples were mounted inside a vacuum chamber of a $^3$He/$^4$He dilution refrigerator. A mechanical feed-through with gears allowed to establish PCs by letting an etched tip of polycrystalline Cu wire (Ø≈0.25 mm) approach and contact the surface of a Zn single crystal. A DC current $I$ with a small superposed AC component $dI$ passed through the contact and the voltage drop $V + dV$ across the contact was measured to obtain the *I-V* characteristics as well as the differential resistance spectrum *dV/dI(V)* by lock-in technique. The second derivative $d(dV/dI)dV \propto d^2V/dI^2$ was calculated from the *dV/dI(V)* curves. In total 26 different PC's were investigated by measuring the temperature dependence of *dV/dI(V)* from 80-100 mK up to critical temperature $T_c$ and higher in steps 0.05 or 0.1K. 9 PC's were studied by contacting with the Cu needle the cleaved (basal) Zn surface, so that current flow in the PC was expected along the *c*-direction. The other 17 PC's were made by contacting with the Cu tip either the cleaned (in alcohol) or the mechanically polished (with fine sandpaper) side surface of the same single crystal, that is they were established mainly perpendicular to the *c*-axis direction.

Figures 1-3 show the experimental data for 3 representative PCs. The PC in Fig.1 was established on the cleaved Zn surface, that is current should flow mainly along the *c*-direction. For the PCs in Figures 2 and 3, the current flow is expected preferably in the basal *ab*-plane. For all PCs the temperature dependence of *dV/dI(V)* was measured and is shown on the upper panels of each figure. In Figs. 1 and 2 (upper panel), the left inset shows the *dV/dI(V)* dependence on a larger scale above $T_c$ of Zn. The right inset shows the excess current $I_{exc}$, that is difference between the *I-V* curves measured at the lowest temperature and in the normal state above $T_c$. The bottom panel of all figures shows the fitting of *dV/dI(V)* at the lowest temperature by the well-accepted modified BTK model either with the one-gap or with the two-gap approach. The insets in the bottom panels of Fig. 1 and 2 show the numerical derivative *d(dV/dI)dV* or PC EPI spectrum with clear phonon structure, while Fig. 3 (bottom inset) displays *dV/dI(V)* on a large scale, which shows a so called zero-bias anomaly and even non-metallic behavior: diminishing of *dV/dI(V)* with increasing voltage. *dV/dI(V)* of this contact above $T_c$ clearly behaves differently compared to *dV/dI(V)* in Figs. 1&2 (left insets) that is no phonon structure is seen.

The experimental *dV/dI(V)* curves at the lowest temperatures, shown in the bottom panel on Figs. 1-3, were fitted according to the modified BTK model within the single- and the two-gap approximation. This fit includes a scaling parameter *S* that specifies the intensity ratio of the experimental *dV/dI(V)* curve with respect to the theoretical one. In the ideal case *S* = 1. The one-gap fit is not perfect, and it requires an enhanced scaling factor *S* (see PCs #1-3 of Table 1). Such an enhanced *S* could be due to a distribution of gaps or a multi-gap structure (see the discussion in Appendix B of [10]). Indeed, the two-gap approach improves the fit, which can be directly in the graph or by comparing the root-mean-square (rms) deviation between fit and measured curves, and shifts $S$ closer to 1.

Let us turn to the second derivatives that represent the EPI spectra. The intensity of the EPI function $g(\varepsilon) \equiv \alpha^2_{PC} F(\varepsilon)$ was calculated from the second derivative of the *I-V* curve according to [2, 11]:

$$R^{-1} \frac{dR}{dV} = \frac{8ed}{3\hbar v_F} \alpha^2_{PC}(\varepsilon) F(\varepsilon)\big|_{\varepsilon = eV}, \qquad (1)$$

where $R = dV/dI$ is the differential PC resistance, e is the electron charge, $d = (16\rho l/3\pi R)^{1/2}$ is the PC diameter, $v_F$ is the Fermi velocity, $F(\varepsilon)$ is the phonon density of states, and $\alpha(\varepsilon)$ is the matrix element of EPI. Thus, from (1) we get:

$$g(\varepsilon) \equiv \alpha^2_{PC} F = \frac{3}{8} \frac{\hbar v_F}{ed} R^{-1} \frac{dR}{dV} \qquad (2)$$

For the calculations we used $v_F \approx 1.8 \cdot 10^6$ m/s and $\rho l \approx 5 \cdot 10^{-16} \Omega \cdot m^2$ for Zn [12]. Table 1 shows $2g_{max}$, the intensity of the principal phonon peak of Zn at about 8 mV. The factor 2 takes into account that Zn occupies and contributes only half of the PC volume of a hetero-contact with Cu.

**Discussion**

First, we consider that the scatter of the SC gap values (see Table 1) is connected with the anisotropic EPI in Zn. This was already mentioned in the AR study of polycrystalline Zn wires [13], reporting the spreading of $2\Delta/kT_c$ ratios between 2.8 and 3.84. Indeed, the SC gap $\Delta$ for PCs #1&2 in Table 1, established in c-direction, is on average about 150 µeV. This is larger than the mean value of about 135 µeV for PCs #3-6 in Table 1, established along the basal plane. However, the crystallographic orientation of the PC core and, thus, the current flow direction, can deviate from the mechanically defined orientation because we do not know the microscopic structure of the PC. Therefore, in the case of strongly anisotropic Zn, the real crystallographic orientation of the PC can be proved by simultaneous measuring the PC EPI spectrum.

The PC EPI spectrum in Fig.1 (inset of the bottom panel) shows a pronounced phonon peak at around 8 mV. This is characteristic for a PC in c-direction of Zn [9]. In contrast, the PC EPI spectrum in Fig. 2 (inset in the bottom panel) displays in addition a broad peak between 12-14 mV, as observed for PCs in the basal plane of Zn [9]. Moreover, the intensity of the spectrum in Fig. 1 is larger than that in Fig. 2 if one compares $2g_{max}$ for the PCs #1-2 and #3-5 in the Table 1. This also agrees with previous observations [9, 12, 14]. Thus, the different shapes and intensities of the PC EPI spectra in Figs. 1&2 strongly indicate that these two PCs are established in different directions: the first one preferably along the *c*-axis and the second one mainly perpendicular to it.

Let us turn to the results of a more perfect two-gap fit. On the average, the smaller gap $\Delta_1$ around 120-140 μeV dominates, independently of the PC orientation. It corresponds to a ratio $2\Delta_1/kT_c$ between 3.2 and 3.7, that is on the average close to the expected BCS value of 3.52. The larger gap $\Delta_2$ between 160 and 200 μeV implies $2\Delta_2/kT_c$ between 4.2 and 5.2. Such a noticeable difference in the gap values indicates that Zn could be a multiband superconductor.

Indeed, the Fermi surface of zinc contains three sheets: the "cap", the "monster" and the "lens", where the calculated EPI constants $\lambda = \int_0^\infty \alpha_{PC}^2(\varepsilon) F(\varepsilon) d\varepsilon / \varepsilon$ have relative magnitudes of 1.0:1.76:2.30 for cap:monster:lens, respectively [15]. Consequently, those sheets should cause different SC gaps. Since the "cap" sheet has an order of magnitude less "surface area" than the others [15], its contribution should be negligible. Thus, it is reasonable to assume two main gaps in Zn corresponding to the "monster" and the "lens" sheets. Moreover, the ratio of our measured average $2\Delta/kT_c$ of the two gaps 3.45/4.7≈0.73 is close to the ratio of the EPI constants λ of the "monster" and the "lens" sheets 1.76/2.30≈0.77.

Furthermore, the anisotropic SC gap of zinc has been determined by microwave absorption [16]. The authors of [16] found that most of the Fermi surface is associated with a reduced SC gap $2\Delta/kT_c = 3.1\pm0.1$, while the gap near the *c* axis has at least $2\Delta/kT_c= 4.0\pm0.2$. This is in line with the bottom limit of the reduced gap values 3.2 and 4.2 from our two-gap fit for the corresponding directions.

We have investigated 26 PCs, all of which showed clear-cut AR like features. From these 26 PCs, only 10 PCs had resolvable EPI maxima in the second derivative. The remaining PCs had in addition to the characteristic AR spectra a zero-bias dip (possibly due to the superconducting proximity effect), a zero-bias anomaly above $T_c$, or a non-metallic behaviour in the range of phonon energies like that in the bottom-panel inset of Fig. 3. This means that AR is a quite robust phenomenon in our Zn PCs, which takes place even at imperfect PCs without EPI structure in the second derivative. Similar observations were also made in [17] for Pb-Fe hetero-contacts prepared by *e*-beam lithography and evaporation. The reason could be that AR takes place within a distance of about one coherence length from the contact, which is quite large in Zn, about 2 μm [13], while PCs with resistance above 10 Ω have a diameter of less than 15 nm. That is the observed EPI signal stems from a region close to the disturbed and less perfect interface, while AR occurs further away in the more clean and perfect bulk of the superconductor.

The next observation is that intensity of the Zn EPI structure $2g_{max}$ (see Table 1) is close to that of ballistic contacts with elastic mean free path $l_e$ larger than the PC size *d*: ≈ 0.2–0.25 for the c-direction and $\approx 0.1$ perpendicular to the c-direction along the [1120] axis according to [9] (note that g($\varepsilon$)=4G($\varepsilon$) of [9]). At the same time all PCs in Table 1 have barrier strength $Z$ close to 0.55, the value for diffusive or "dirty" PC's according to theory [4, 18]. On the other hand, $Z$ values of about 0.5 were reported recently for PCs with resistances from less than 1 $\Omega$ to above 1 k$\Omega$ between different normal metals and superconducting In [19] and Al [20], respectively. It is unlikely that some permanent tunnel "barrier" with constant thickness exists at the interfaces of all those contacts. Against such a "barrier" speak also the results of [17], where $dV/dI$ of Pb-Fe PCs prepared by *e*-beam lithography and oxide-free Pb and Fe evaporation, appeared to have a similar "barrier strength". Again $l_e$ must be compared with the coherence length in Zn, but not with the PC size.

Finally, we turn to the excess current $I_{exc}$, which is shown in the right insert in the upper panel of each figure. $I_{exc}$ reaches its maximum value at about 1 mV, that is at e$V$>>$\Delta$ as theory says, with its normalized value e$I_{exc}R/\Delta$ of about 0.6 close to the theoretical 0.73 for "dirty" or diffusive superconductor-normal metal contacts [4]. If the fit of *dV/dI(V)* would be perfect with $S$=1 and $Z \approx 0.55$, then e$I_{exc}R/\Delta$ should equal the theoretical value. A smaller Z would enhance it even further. We attribute the reduction of our experimental e$I_{exc}R/\Delta$ with respect to theory to Dynes' lifetime parameter $\Gamma$. Like the Z parameter, the e$I_{exc}R/\Delta$ product does not depend on whether the PC spectrum shows EPI maxima or not.

**Conclusions**

Our study of a Zn single crystal by applying *simultaneously* Andreev reflection and Yanson point-contact spectroscopy resulted in the following:

1. We have found a correlation between the anisotropic EPI function and the SC gap in Zn. The average SC gap is about 10% larger for the c-direction than in the basal plane, which coincides with the larger EPI intensity for the c-direction.
2. There is evidence for multiband superconductivity in Zn with two main gaps with the reduced gap ratio 2$\Delta/kT_c$ between 3.2 and 3.7 for the small gap and between 4.2 and 5.2 for the larger one. We attribute the smaller gap to the "monster" and the larger gap to the "lens" sheet of the Fermi surface of Zn.
3. AR features are more robust with respect to the PC quality than the EPI ones, possibly due to the much larger coherence length in Zn compared to the size of a PC core.
4. For PCs close to the ballistic regime with intense EPI spectra the "barrier strength" $Z$ evaluated from the AR spectra is near the value predicted for the diffusive regime of current flow in superconducting PCs. It looks like the characteristic length scale for diffusive transport differs for the two mechanisms: it is the elastic electron mean free path $l_e$ relative to the PC size for the EPI features, and $l_e$ relative to the coherence length for AR.

**Acknowledgements**

YGN is very grateful to the Department of Physics and Astronomy, University of Turku, where all measurements have been done, for hospitality. We thank N. L. Bobrov and L.V. Tiutrina for letting us use their two-gap fit program for the modified BTK model, E. Tuuli for helping us handling the dilution refrigerator, and the Jenny and Antti Wihuri Foundation for financial support.

Table 1. Parameters of the spectra evaluated for selected PCs with EPI features and not disturbed by proximity-effect induced AR structures. The contacts have normal-state resistance $R_{PC}$. Along with the SC gap $\Delta$, the broadening $\Gamma$ (or Dynes' lifetime) parameter and "barrier" parameter $Z$, we show also the scaling parameter $S$, the weight factor of the first gap $w$ ($w \leq 1$), the intensity of the first Zn phonon peak $g_{max}$ and the reduced gap ratio $2\Delta/kT_c$. Note that the critical temperature of the contact $T_c$ was estimated with an accuracy of a few tenths of K due to stepwise measurement of the $dV/dI$ spectra. It also turned out that average gap $<\Delta> = w\Delta_1 + (1-w)\Delta_2$ (shown in Figs.1-3) nearly equal the gap $\Delta$ of the one-gap fit. PCs 1, 4 and 6 are shown in Figs. 1, 2 & 3, respectively.

| | | | 1 gap fit | | | | | 2 gap fit | | | | | | |
|---|---|---|---|---|---|---|---|---|---|---|---|---|---|---|
| | # | $R_{PC}$ $\Omega$ | $T_c$ K | $\Delta$ µeV | $\Gamma$ µeV | $Z$ | $S$ | $\Delta_{1,2}$ µeV | $\Gamma_{1,2}$ µeV | $Z$ | $S$ | $w$ % | $2g_{max}$ | $\frac{2\Delta}{kT_c}$ |
| c-dir | 1 | **43** | 0.87 | 154 | 44 | 0.47 | 1.23 | 135/198 | 35/23 | 0.46 | 1.09 | 74 | 0.26 | 4.0 |
| | 2 | 13 | 0.88 | 145 | 30 | 0.52 | 1.3 | 125/165 | 17/15 | 0.51 | 1.07 | 51 | 0.26 | 3.8 |
| ab-direction | 3 | 18 | 0.87 | 130 | 29 | 0.52 | 1.22 | 119/161 | 18/10 | 0.51 | 1.03 | 77 | 0.15 | 3.5 |
| | 4 | **9.5** | 0.87 | 128 | 14 | 0.53 | 1.06 | 118/152 | 4/10 | 0.51 | 0.94 | 61 | 0.15 | 3.4 |
| | 5 | 22 | 0.86 | 147 | 11 | 0.44 | 1.01 | 140/178 | 4/8 | 0.43 | 0.94 | 78 | 0.2 | 4.0 |
| | 6 | **180** | 0.85 | 130 | 48 | 0.55 | 1.45 | 130/177 | 45/45 | 0.55 | 1.36 | 99 | – | 3.5 |

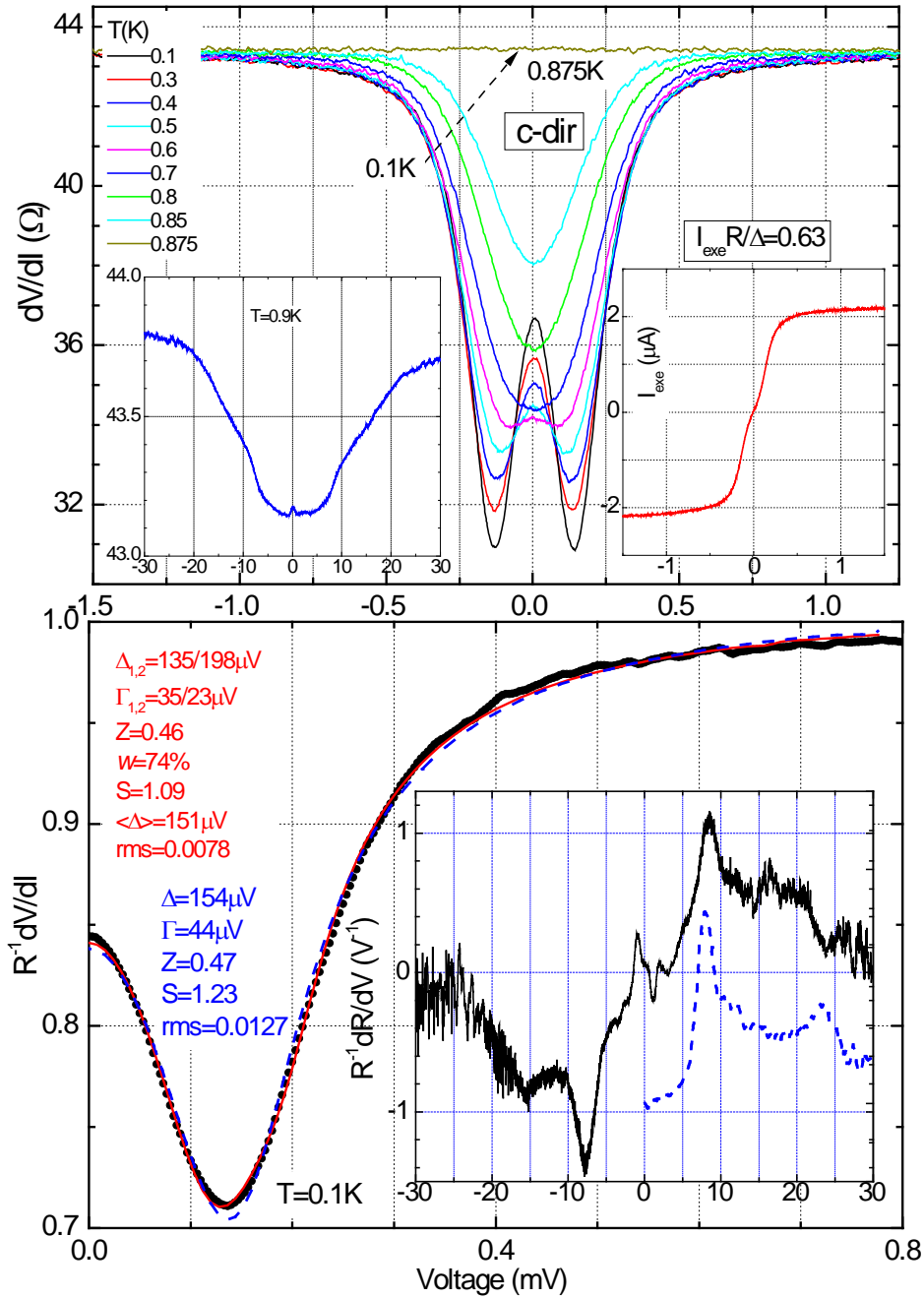

Fig. 1. Upper panel: *dV/dI(V)* for a PC along the c-axis taken at different temperatures from the lowest one up to above $T_c$. Left inset: *dV/dI(V)* above $T_c$ for the larger voltages. Right inset: excess current $I_{\text{exc}}$. Bottom panel: experimental *dV/dI(V)* at the lowest temperature 0.1K (symbols) along with fit by the one-gap (dashed blue curve) and two-gap (solid red curve) model. Also shown are the parameters of one- and two-gap fit (see Table 1 caption for the definition, rms is root mean square error). Inset: numerical derivative of *dV/dI(V)* from the left insert in the upper panel along with PC EPI spectrum of Zn along c-axis (bottom blue curve) from [14] for comparison. The enhanced weight of the derivative between 15 and 20 mV could be due to the copper contribution, which has an intense and broad transverse phonon peak in this region.

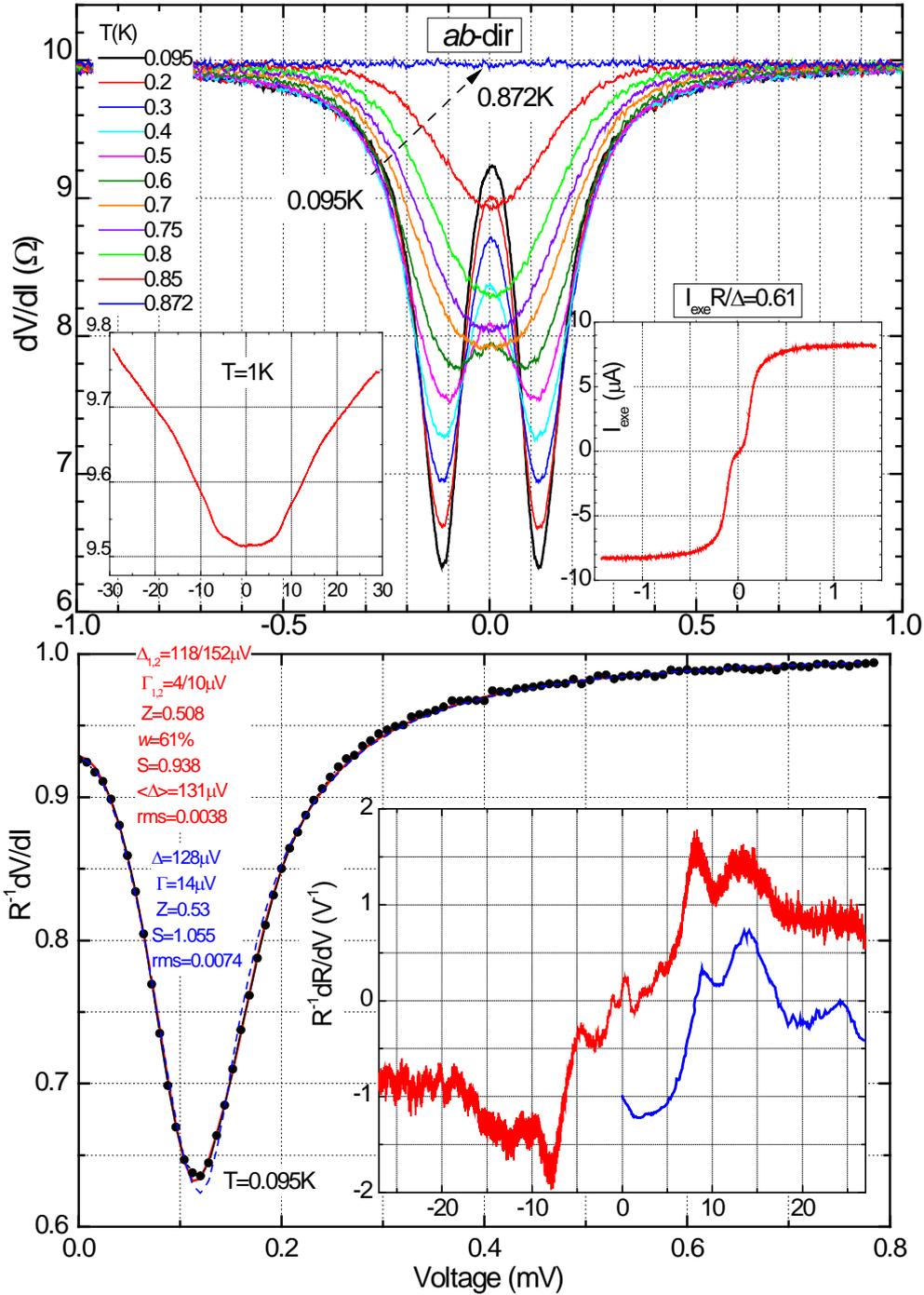

Fig. 2. Upper panel: *dV/dI(V)* for a PC in the basal plane taken at different temperatures from the lowest one up to above $T_c$. Left inset: *dV/dI(V)* above $T_c$ for larger voltages. Right inset: excess current $I_{exc}$. Bottom panel: experimental *dV/dI(V)* at the lowest temperature 0.095K (symbols) along with the fit by the one-gap (dashed blue curve) and the two-gap (solid red curve) model. Also shown are the parameters of one- and the two-gap fit (see Table caption for the definition, rms is root mean square error). Inset: numerical derivative of *dV/dI(V)* from the left insert in the upper panel along with PC EPI spectrum of Zn along the basal plane (bottom blue curve) from [14] for comparison.

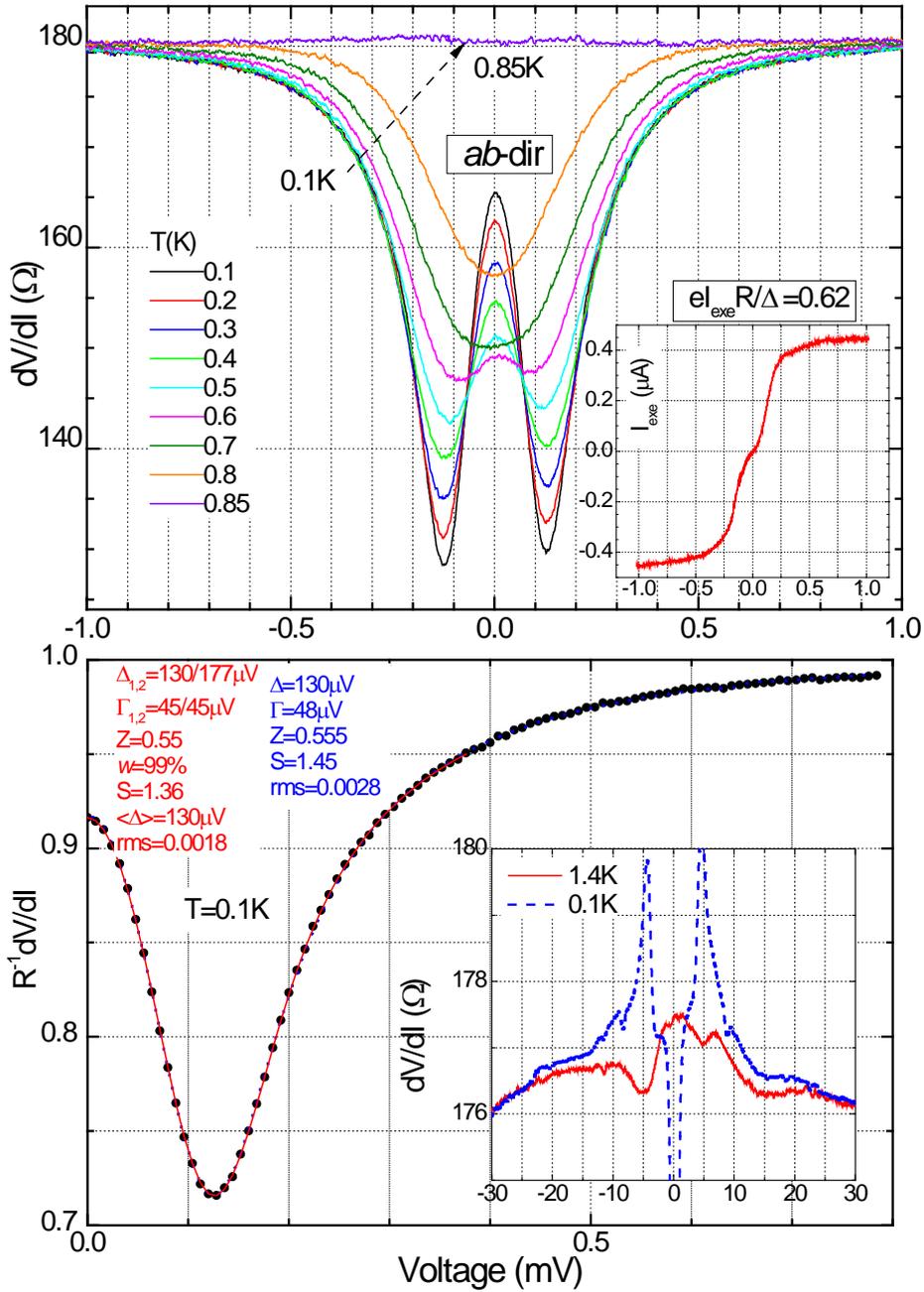

Fig. 3. Upper panel: *dV/dI(V)* for a PC in the basal plane taken at different temperatures from the lowest one up to above T$_c$. Inset: excess current *I*$_{exe}$. Bottom panel: experimental *dV/dI(V)* at the lowest temperature 0.1K (symbols) along with a fit by one-gap (dashed blue curve) and the two-gap (solid red curve) model (both fits are almost indistinguishable). The two-gap fit results in either two gaps very close to each other or if the second gap differs strongly from the first one, it contributes less than 1%. Also shown are the parameters of one- and two-gap fit (see Table 1 caption for the definition, rms is root mean square error). Inset: *dV/dI(V)* above (solid blue curve) and below T$_c$ (dashed red curve) for the larger voltages. Note the zero-bias anomaly (maximum) of the curve above T$_c$ and the "side" peaks of the curve below T$_c$.